\documentstyle[12pt,aasms4]{article}
\def\chandra{{\it Chandra}}

\def\hst{{\it HST}}

\def\rosat{{\it ROSAT}} 
\def\ein{{\it Einstein}} 
 
\def\merlin{{\it MERLIN}}

\def\nh{cm$^{-2}$}
\def\arcsec{$^{\prime\prime}$}

\def\ltsima{$\; \buildrel < \over \sim \;$}
\def\simlt{\lower.5ex\hbox{\ltsima}} % < over ~
\def\gtsima{$\; \buildrel > \over \sim \;$}
\def\simgt{\lower.5ex\hbox{\gtsima}} % > over ~

\begin{document}

\title{Chandra Observations of the X-Ray Jet of 3C~273}

\author{Rita M. Sambruna} \affil{George Mason University, Dept. of
Physics \& Astronomy and School of Computational Sciences, 4400
University Drive, M/S 3F3, Fairfax, VA 22030}

\author{C. Megan Urry}
\affil{STScI, 3700 San Martin Dr., Baltimore 21218, MD}

\author{F. Tavecchio and L. Maraschi}
\affil{Osservatorio Astronomico di Brera, via Brera 28, 20121 Milano, Italy} 

\author{R. Scarpa} 
\affil{STScI, 3700 San Martin Dr., Baltimore 21218, MD}

\author{G. Chartas}
\affil{The Pennsylvania State University, Department of Astronomy and
Astrophysics, 525 Davey Lab, State College, PA 16802}

\author{T. Muxlow}
\affil{Jodrell Bank Observatory, University of Manchester,
Macclesfield, Cheshire, SK11 9DL, UK}

\begin{abstract}

We report results from \chandra\ observations of the X-ray jet of
3C~273 during the calibration phase in 2000 January. The zeroeth-order
images and spectra from two 40-ks exposures with the HETG and
LETG+ACIS-S show a complex X-ray structure. The brightest optical
knots are detected and resolved in the 0.2--8~keV energy band.  The
X-ray morphology tracks well the optical. However, while the X-ray
brightness decreases along the jet, the outer parts of the jet tend to
be increasingly bright with increasing wavelength. The spectral energy
distributions of four selected regions can best be explained by
inverse Compton scattering of (beamed) cosmic microwave background
photons. The model parameters are compatible with equipartition and a
moderate Doppler factor, which is consistent with the one-sidedness of
the jet. Alternative models either imply implausible physical
conditions and energetics (the synchrotron self-Compton model) or are
sufficiently ad hoc to be unconstrained by the present data
(synchrotron radiation from a spatially or temporally distinct
particle population).

\end{abstract} 

{\sl Subject Headings:}{
galaxies: active --- 
galaxies: jets --- 
(galaxies:) quasars: individual (3C273) ---
X-rays: galaxies} 

\section{Introduction} 

%The discovery of the X-ray emitting jet in the luminous quasar
%PKS~0637--752 ($z=0.651$) has renewed attention to the physics of
%resolved, kiloparsec-scale jets. Because the jet of 3C~273 is more
%than four times closer ($z=0.158$), its X-ray emission is brighter and
%can be resolved on smaller scales. Thus this jet is ideal for
%intensive multiwavelength study.

Being nearby ($z=0.158$), the jet of 3C~273 is ideal for intensive
multiwavelength studies. The 10\arcsec-long radio jet has a knotty
morphology, with the first detected knot at $\sim 13$\arcsec\ from the
core and increasing radio intensity toward a terminal bright hot spot
$\sim22$\arcsec\ from the core (Flatters \& Conway 1985). Variable
radio polarization and low Faraday rotation have been observed along
the jet (Conway et al. 1993). In the near-infrared, the jet brightness
tracks well the radio, while at optical wavelengths the jet is much
narrower and concentrated along the axis (R\"oser \& Meisenheimer
1991; Bahcall et al. 1995; Neumann, Meiseinheimer, \& R\"oser
1997). This is globally consistent with synchrotron emission being the
dominant mechanism from radio through optical.

On much smaller (VLBI) scales, 3C~273 has a radio jet with
superluminal components (Davis, Muxlow, \& Unwin 1991), implying bulk
relativistic motion of the emitting plasma, with a jet-to-counterjet
ratio of 5300.  The resulting Doppler beaming provides a natural
explanation for the absence of a visible counterjet.  On kiloparsec
scales, the X-ray/optical emission remains asymmetric, suggesting
%at least $10^4$ in the
%optical (derived by us from archival \hst\ images), suggesting
relativistic bulk motion may still be present on large scales.

In the X-rays, the 3C~273 jet was first detected with \ein\ and
\rosat\ (Harris \& Stern 1987; R\"oser et al. 2000), with most of the
X-ray flux coming from the brightest optical knot. The spectral energy
distributions from radio to X-rays of the individual knots, resolved
at the \rosat\ HRI resolution ($\sim15$\arcsec), show a variety of
shapes. For the brightest knot, the X-ray flux appeared to lie on the
extrapolation from the radio-optical spectrum, suggesting synchrotron
emission could extend to at least 2~keV. For the remaining knots, the
X-rays were above the extrapolation from the longer wavelengths
(R\"oser et al. 2000).

Here we use the \chandra\ ACIS-S observation of the 3C~273 jet
obtained during the calibration phase (80~ks exposure total) to study
the jet morphology in greater detail. The \chandra\ data are compared
to observations at longer wavelengths using archival \hst\ U- and
R-band data and deep \merlin\ observations at 1.65~GHz.
%\footnote{Based on observations made with the NASA/ESA
%Hubble Space Telescope, obtained at the Space Telescope Science
%Institute, which is operated by the Association of Universities for
%Research in Astronomy, Inc., under NASA contract NAS~5-26555.} and
%In \S~2 we describe the
%observations and data analysis and in \S~3 we describe the
%multiwavelength morphology of the jet. In \S~4 we test plausible
%models for the radio through X-ray emission and in \S~5 we give our
%conclusions. 
Throughout this work we adopt H$_0=75$ km/s/Mpc and $q_0=0.5$, so at
the distance of 3C~273, 1\arcsec=2.4~kpc. 

\section{Observations and Data Analysis} 

\chandra\ observed 3C~273 during the calibration phase in January
2000, with ACIS-S (Garmire et al. 2000) plus the Low Energy
Transmission Grating (LETG) for 40~ks on January~9--10, and with
ACIS-S plus the High Energy Transmission Grating (HETG) for 40~ks on
January~10. Here we concentrate on the zeroeth-order (undispersed)
HETG and LETG images of the jet. After screening out data
corresponding to bad aspect times and grades, 39.8~ks were left in the
LETG and 39.4~ks in the HETG exposures, which were gain corrected
using the appropriate files for the dates of observation.  The S3
background was stable during both observations.  Images were extracted
in the energy range 0.2--8~keV, where the background is lower ($\sim
8.6 \times 10^{-6}$ c s$^{-1}$ arcsec$^{-2}$ for LETG, $\sim 5.1
\times 10^{-6}$ c s$^{-1}$ arcsec$^{-2}$ for HETG). We also
accumulated separately soft (0.2--1~keV) and hard (1--8~keV) images
calibrated in flux, in order to study spectral evolution along the
jet. 

To maximize the spatial resolution in the final image, we deconvolved
the data using the maximum likelihood algorithm, following Chartas et
al. (2000). The HETG and LETG images were deconvolved separately, then
stacked together. The effective resolution (FWHM) of the deconvolved
X-ray images is 0.37\arcsec\ FWHM. However, fluxes were extracted from
the original, un-deconvolved event files. 
% making no correction for their somewhat different effective areas.

For X-ray spectral analysis only the ACIS-S HETG zeroeth-order image
was used since the ACIS+LETG configuration is not as well calibrated
yet. Position-dependent responses and ancillary files were created,
and spectra were rebinned to have at least 20 counts per bin, to
validate the use of the $\chi^2$ statistic. 
%Good agreement between the
%LETG and HETG fluxes was found at hard energies, while at energies
%$\lesssim 1$~keV the flux from the LETG data was higher by a factor
%2--4 than from the HETG data, implying one or both of the instrument
%calibrations is not yet correct. 
% MEG I TOOK THIS SENTENCE OUT BECAUSE IT IS TOO PROVOCATIVE 
Analysis of calibration HETG observations of Capella and simulations
with \verb+MARX+ v.3.0 show that the contribution of the core PSF
wings is negligible (less than 1 count/bin) at $\sim$ 13\arcsec\ or
more from the core, where the first bright jet knot is
detected. Because of the roll angle, the jet X-ray emission is
unaffected by the grating diffraction spikes in both the HETG and LETG
data. 

\hst\ images of 3C~273 obtained in the R and U bands (first reported
by R\"oser et al. 1997) and \merlin\ data at 1.65 GHz (first reported
by Bahcall et al. 1995) were reanalyzed in order to extract spectra in
apertures matched to those used in the X-rays. The resolution of the
\merlin\ data consists of a beamwidth 0.18\arcsec x 0.14\arcsec\ FWHM at
a position angle of 27 deg (Bahcall et al. 1995). 
%Four and ten \hst\ images of $\sim 2600$ s each were available for
%filter R and U, respectively.  Images were aligned within 0.2 pixels,
%and combined using the IRAF task CRREJ, in order to obtain a final,
%cosmic-ray free image. To isolate the emission from the jet, the sky
%background, as well as diffuse light from the bright, over-exposed
%nucleus was modeled with a surface and then subtracted from the image.
%The final flux calibration is based on the photometric zero points
%given by Holtzmann et al. (1995).  The \merlin\ image is basically
%that which appears in Bahcall et al, but rotated and regridded by the
%AIPS task LGEOM.  
The final radio-optical-X-ray spectra of the jet knots are therefore
accurate representations of the spectra of the X-ray-emitting regions.
%In the optical, one-dimensional surface brightness profiles along the
%jet were extracted by collapsing the emission to a central line. In
The radial profiles at all wavelengths were extracted by running a
1\arcsec\ slice orthogonally to the jet axis.  The resolution of the
radio and optical data is 0.16\arcsec/pixel and 0.046\arcsec/pixel,
respectively.

\section{Results}

\subsection{X-Ray, Optical, and Radio Morphologies of Jet}

Figure~1 shows the \chandra\ ACIS-S image of the 3C~273 jet in the
0.2--8~keV band.  The X-ray jet is $\sim$8\arcsec-long and has a
knotty morphology, starting with a bright, resolved knot at $\sim
13$\arcsec\ from the core, coincident with optical knot A1 (using
notation from Bahcall et al. 1995). This is followed by a second
bright knot at $\sim 15$\arcsec, coincident with optical knot B1, with
weak X-ray emission in between. The X-ray emission then fades to a low
but still detectable level before terminating at knot D. No X-ray
emission was detected between the nucleus and knot A1. A total of 1361
counts were detected from the jet in the 0.2--8~keV band (614 from the
HETG, 747 from the LETG), of which 584 are contained in the first knot
A1 within 0.9\arcsec.

Figures~2a--d shows the jet surface brightness profiles at X-ray,
optical (U and R bands) and radio (1.65 GHz) wavelengths, normalized
to the peak intensity in each band. Nearly all the optical and radio
knots are detected at X-rays, the exception being the final hot spot
(called ``H'' by Bahcall et al. 1995) at $\sim 21.5$\arcsec, which is
bright in the radio and optical but has no obvious X-ray
counterpart. The peaks align to within a few tenths of an arcsec, well
within the astrometric and attitude accuracies of \chandra\ and \hst.
The four normalized profiles show that, while the X-ray brightness
decreases along the jet, at longer wavelengths the outer parts of the
jet become increasingly dominant. Note that this behaviour appears
rather regular through the U and R bands down to the radio.

% X-ray and radio emission is generally anti-correlated, while the
%optical is intermediate.

%The profiles in Figure~2 were normalized to the maximum observed
%intensity in order to show the relative scaling of the flux in the
%various knots at the four wavelengths. A general anti-correlation is
%apparent between X-rays and radio, with the X-ray brightness
%decreasing from the inner to the outer regions, while the radio
%brightens. The optical appears somewhat intermediate, being more
%nearly uniform along the jet and with strongest emission at knots A
%(particularly in the U band) and D (in the R band).

\subsection{Spectral Energy Distributions} 
 
We constructed radio-to-X-ray Spectral Energy Distributions (SEDs) for
four interesting regions along the jet, identified in the \chandra\
profile in Figure~2 as A, B, C, and D. Fluxes at X-ray, optical, and
radio wavelengths were extracted from circles positioned at the
centers of these regions and with radii equal to half the marked
size. Region A corresponds to knot A1 (maximum X-ray flux); region B
encloses optical knots A2, B1, and B2; region C includes knots C1--C3;
and region D contains knots D1--D3. The X-ray fluxes were corrected
for absorption using a column density equal to the Galactic value,
N$_H=1.8 \times 10^{20}$ \nh\ (Elvis et al. 1989), which is consistent
with the X-ray spectrum of knot A1 and the optical fluxes were
dereddened assuming A$_V$=0.1.

Figure~3 shows the SEDs of the four regions of the jet. In all cases,
the X-ray flux lies above the extrapolation from the radio-to-optical
continuum, which steepens in the optical band.  Thus the X-rays appear
to belong to a distinct spectral component. The radio-to-optical flux
ratio increases from the inner to the outer parts, in accordance with
the steepening of the optical continuum (e.g., R\"oser \& Meisenheimer
1991).  The X-ray spectrum of region A in the energy range 2--8~keV
was fitted by a power law with photon index 
$\Gamma_x=2.1^{+0.5}_{-0.3}$ (90\% confidence interval) and Galactic
absorption.
%(N$_H=1.8 \times 10^{20}$ \nh, Elvis et al. 1989).
%Also plotted is the X-ray spectrum of region A, a relatively flat power law.
Region B has a similar X-ray spectrum but larger errors due to the
lower flux, and there are too few photons for spectral analysis in the
remaining regions. We can not set constraints on the low-energy
($\lesssim 1$~keV) spectrum given the current calibration
uncertainties.

From the hardness ratio, HR=1--8~keV/0.2--1~keV, there is a slight
indication of spectral softening along the jet, but only at the 2
sigma level: HR=1.2 $\pm$ 0.1 for region A and HR=0.9 $\pm$ 0.2 for
region D (using uncertainties as prescribed by Gehrels 1986). With
better statistics, we could determine whether this effect is real.
% If so, it would be extremely interesting because it would indicate
%spectral aging of the radiating electrons.

\section{Discussion: The Multiwavelength Jet Spectra} 

%The high
%radio intensity at the end of the jet probably comes from an increase
%in the magnetic field due to compression at the working surface of the
%jet. The jet likely decelerates in this region, and the electrons cool
%faster.

While the radio-through-optical spectrum from the 3C~273 jet is almost
certainly due to synchrotron radiation (see \S~1), the X-rays can not
be attributed to synchrotron emission from the same population of
electrons, since this would be inconsistent with the shape of the SEDs
in Figure~3, except possibly for region A (and with the different jet
morphologies in Figure~2). We conclude that in all four regions the
X-rays arise from a different spectral component than does the
radio-optical emission.

We considered the following possibilities for the origin of the
X-rays: (1) synchrotron emission from a second, much more energetic
population of electrons (R\"oser et al. 2000); (2) inverse Compton
scattering of radio-optical synchrotron photons; (3) inverse Compton
scattering of photons external to the jet. We can rule out a fourth
possibility, that the 0.8--2~keV X-rays are bremmstrahlung from
thermal gas confining the radio-optical jet, since the required
density is higher than allowed by the observed (low) rotation measure
(Conway et al. 1993); such a model is also inconsistent with the X-ray
spectrum of knot A and with the general morphological correspondence
of X-ray and radio-optical (synchrotron) emission. The present data do
not constrain the presence of thermal gas cooler than $\sim1$~keV.

\subsection{Synchrotron Emission from Two Populations of Electrons}

If the second electron population is co-spatial with the first, it
must arise from a separate episode of acceleration (since cooling
would create a spectrum extending smoothly to lower energies). Its
mean electron energy must also be higher (for the same magnetic
field), and the acceleration event must have occurred more recently
(the electrons not yet having cooled substantially).  One possibility
is that hotter electrons are produced via proton-induced cascades
(Mannheim \& Biermann 1992); this model is not constrained by the
limited spectral data discussed here.

A simple homogeneous synchrotron model fits the radio-optical emission
of the four regions with plausible parameters: $B$ between 1 and 10
$\mu$Gauss, Doppler factor $\delta\sim5$, size $\sim 5 \times
10^{21}$~cm, and electron cutoff energies $\gamma_{\rm min} \sim 10$
and $\gamma_{\rm max} \sim 10^6$. For the same magnetic field,
X-ray-emitting electrons would have to have $\gamma_{\rm min}$ in the
range $10^{6-7}$ while $\gamma_{\rm max}$ is unconstrained as we do
not know the upper-cutoff frequency of the spectrum. 
%(Note that
% future sensitive gamma-ray instruments could set constraints 
% observed MeV or
%GeV emission from the 3C~273 nucleus; von Montigny et al. 1997) 
The energy distribution of this second population would have a shape
similar to the radio-optical-emitting electrons (index 2.6).
Alternatively, the second electron population could occupy a distinct
region in the jet, with a different magnetic field, in which case
there are enough free parameters that the physical state of the jet is
essentially not constrained by the present observations.

\subsection{Synchrotron Self-Compton (SSC) Emission}

X-rays are inevitably produced by synchrotron self-Compton emission,
meaning inverse Compton scattering of the radio-optical synchrotron
photons by the electrons that emit them. This model can be
significantly constrained by a few observed fluxes, at least in the
homogeneous approximation (Tavecchio, Maraschi \& Ghisellini 1998).
However, the only way for the Compton component to be more luminous
than the synchrotron component (as is the case in regions A--C;
Fig.~3) is if the observed emission is weakly beamed or de-beamed
($\delta \lesssim 1$) which enhances the estimate of the intrinsic
photon density.  For $\delta \simeq 1$ one needs additionally a very
low magnetic field ($B\lesssim10^{-6}$~G), orders of magnitude below
the equipartition value (see the discussion for the case of
PKS~0637--752; Tavecchio et al. 2000).  A more plausible value of the
magnetic field can be recovered if the jet is significantly debeamed
($B\simeq 10^{-5} - 10^{-4}$~G for $\delta \simeq 0.5-0.3$). This
implies much larger total jet power and is energetically less
favorable. It also implies the jet is misaligned, which is implausible
given
%the strong limits on a counterjet and
the blazar-like core radio source. Instead, if applicable on large
scale, the jet-to-counterjet ratio of $5300$ implies a bulk Lorentz
factor $\Gamma_{\rm bulk} > 3$ and viewing angle $\theta < 18$, or
$\delta > 6$ (for $\alpha=0.5$).
%
%In this model, the X-ray spectrum should have the same slope as the
%radio spectrum since the same low-energy electrons ($\gamma \sim
%100$) are involved; fitting a synchrotron self-Compton model to 
%the spectrum of region A implies a somewhat steeper slope than 
%the observed X-ray value ($\alpha_x \sim 1.1$) and
%published radio spectra have $\alpha_r \lesssim 0.9$ 
%(Conway et al. 1993), at least at knot A. 
%RITA, I DELETED THE ABOVE BECAUSE IT IS ACTUALLY INCONSISTENT. EITHER
%THE X-RAYS ARE TOO STEEP (STEEPER THAN THE RADIO) OR TOO FLAT (FLATTER
%THAN THE SSC MODEL FIT). SO WE REALLY LEARN NOTHING FROM THE SPECTRUM, 
%SO TAKE IT OUT.
Thus, while SSC emission must be produced, for plausible assumptions it
is insufficient to explain the observed X-rays.

\subsection{External Compton Emission}

X-rays can be produced when relativistic electrons in the jet inverse
Compton scatter photons external to the jet (external Compton, or
EC). Because the jet is far from the nuclear region, however, the
energy density of ambient photons is negligible unless the jet is
still relativistic on kiloparsec scales. In that case, the (forward)
cosmic microwave background (CMB) photons are beamed in the jet frame,
so their intrinsically low rest-frame energy density, $7 \times
10^{-13}$~ergs~s$^{-1}$, is enhanced by the factor $\Gamma_{\rm
bulk}^2$ (Tavecchio et al. 2000).
%Celotti et al. 2000). 

The EC/CMB model is as fully specified as the SSC model, so similar
constraints on $B$ and $\delta$ apply (Tavecchio et al. 2000). In this
case, however, the model needs a significant beaming factor
($\delta\sim5$) and can be reasonably close to an equipartition of
energy between magnetic field and electrons ($B\sim 10^{-5}$~G).  A
model SED for knot A is shown in Fig. 3.

The observed decreasing trend of the X-ray/radio luminosity ratio
along the jet finds no obvious explanation in any of the scenarios
discussed above.  Ad hoc variations of the involved parameters must be
invoked.
%and the lack of X-ray flux at the end of the jet 
Within the EC/CMB scenario the ratio of magnetic to photon energy
density in the jet can increase because either the magnetic field
increases or the amplification factor for the CMB energy density as
seen in the jet ($\Gamma_{\rm bulk}^2$) decreases, or both. In Fig. 3
we show a model SED for knot C, obtained by increasing only the
magnetic field by a factor $\simeq 4$ with respect to knot A together
with a model SED for knot D obtained increasing $B$ by a further
factor of 2 and decreasing $\Gamma$ by a factor $\simeq$ 4 which
leaves $\delta$ almost constant. The precise values of the parameters
used are given in the figure Caption.  An enhancement of the magnetic
field by compression, as the plasma in the jet goes through successive
shocks, is plausible as indicated by polarization maps (Conway et
al. 1993). Deceleration at shocks is also plausible just before the
main radio hot spot, at which deceleration is presumably so strong
(e.g., R\"oser, Conway, \& Meisenheimer 1996), as to drastically
suppress X-rays from inverse Compton scattering of CMB photons.

In the EC/CMB model the X-ray emission is produced by electrons of
relatively low energy ($\gamma\sim1000$) (whose spectrum is measurable
in the radio). Therefore the cooling timescale is long, implying that
a spctral steepening along the jet can not be due to cooling. 

Electron cooling also cannot explain the observed knottiness of the
X-ray jet which could be due to compression of particles and fields
associated with shocks. Alternatively the front of the relativistic
plasma may describe a helix around the jet and thus have different
angles with respect to the line of sight (even if fixed Lorentz
factor). This is consistent with the optical morphology in the \hst\
images, which show a ``corkscrew'' jet (Bahcall et al. 1995). The
X-rays would then be brightest where the alignment was favorable. The
physics derived here would be unchanged.

While in our model X-ray emitting electrons have long cooling times,
electrons emitting synchrotron optical radiation ( close to
$\gamma_{max}$) cool in a time short compared to the travel time along
the jet; the cooling time at knot A is $\sim 5\times 10^{10}$ s,
corresponding to a distance of less then a few kpc. These electrons
must therefore be accelerated in situ.  From our spectral fits the
maximum electron Lorentz factor $\gamma_{max}$ appears to decrease
along the jet (see parameters in Fig. 3 caption). This could be
understood as an overall cooling of the accelerated electrons or as a
decrease in the efficiency of shock acceleration or a combination of
both.

\section{Conclusions} 

We have analyzed the multiwavelength jet in 3C~273, in an attempt to
determine its physical state. Of the possible mechanisms for the X-ray
emission, the most plausible is inverse Compton scattering of
microwave background photons (EC/CMB), which are seen as beamed by jet
electrons if there is bulk relativistic motion on kiloparsec
scales. The contribution from SSC X-rays is much smaller unless the
magnetic field is well below equipartition and/or the observed
emission is de-beamed, either of which increases the required jet
power to uncomfortable levels. An alternative picture, in which the
X-rays come from direct synchrotron radiation, posits a second,
distinct population of electrons, which might be produced in a second
region (unresolved by the X-ray data) more recently than the cooler
population responsible for the radio-optical spectrum. This is a
somewhat {\it ad hoc} model which can not be well constrained given
the large number of free parameters. Future longer \chandra\
observations of the 3C~273 jet could provide more sensitive
constraints both spatially and spectrally.

\acknowledgements

This work was partially supported by NASA grants NAG5--7925 (RMS),
NAG5--9327 (CMU, FT, LM), the Italian MURST (MURST-COFIN-98-02-15-41)
and ASI (ASI-ARS-98-74 and ASI-I/R/105/00) and from the European
Commission (ERBFMRX-CT98-0195).  Support was also provided by NASA
through grants GO-06363.01-95A and GO-07893.01-96A from STScI,
operated by AURA, Inc., under NASA contract NAS~5-26555.  \merlin\ is
a national facility operated by the University of Manchester on behalf
of PPARC in the UK.

%\newpage 

%\vspace{1cm}

\newpage 

\noindent{\bf Figure~1:} \chandra\ ACIS-S image of the jet of 3C~273
(color scale with superimposed contours), from combining two exposures
obtained during the calibration phase in 2000 January, for a total
exposure of 79.2~ks. A maximum likelihood algorithm was applied to the
stacked image (Chartas et al. 2000), yielding a pixel size of
0.125\arcsec/pixel and an effective resolution at FWHM of
0.37\arcsec. A total of 1361 counts were detected from 0.2--8~keV.
The nucleus of 3C~273 is 13\arcsec\ from this knot, well outside the
figure. 

\noindent{\bf Figure~2:} Multiwavelength surface brightness profiles
along the jet, normalized to the brightest peak in each band. X-ray
data from \chandra, U- and R-band data from archival \hst\ images, and
radio data at 1.65~GHz from \merlin\ observations. The dashed line
in the bottom panel show the radio profile multiplied by 10, in order
to show the jet structure at radii \simlt 20\arcsec. 
%The X-ray
%morphology conforms closely to the optical morphology. The X-rays are
%brightest at the beginning of the jet, while the radio peaks at the
%end and the optical jet is intermediate. 
Regions extracted for subsequent spectral analysis are marked A--D at
the top. The normalization factors are: 87.2734 counts/bin for the
X-rays, 0.17 $\mu$Jy/pixel for the U band, 0.38 $\mu$Jy/pixel for the
R band, and 0.37 Jy/pixel for the radio.

\noindent{\bf Figure~3:} Spectral energy distributions for regions A,
B, C, and D (defined in Fig.~2), extracted from the \merlin, \hst, and
\chandra\ images using matching apertures fixed by the X-ray
resolution. The uncertainties on the X-ray fluxes are 40\%, while the
optical and radio data have 10\% uncertainties. 
%For all four regions,
%the X-rays lie above any smooth extrapolation of the radio-optical
%spectrum, ruling out synchrotron emission from a single population of
%electrons. The spectra steepen continuously moving outward along the
%jet from region A to region D. 
The solid lines represent fits to the spectra with the EC/CMB model
(see text), with the following parameters: {\it A:} minimum and
maximum electron energy $\gamma_{min}=20$, $\gamma_{max}=6\times10^6$,
index $n=2.6$, density $K=8.1\times10^{-3}$ cm$^{-3}$, magnetic field
$B=1.9\times 10^{-6}$ G, region size $R=5\times10^{21}$ cm, Doppler
factor $\delta=5.2$; {\it B:} like A, except
$\gamma_{max}=3\times10^6$ and $B=3.6\times 10^{-6}$ G; {\it C:} like
A, except $\gamma_{max}=1.5\times10^6$ and $B=8.8\times 10^{-6}$ G;
{\it D:} like A, except $\gamma_{max}=7\times10^5$, $B=2.2\times
10^{-5}$ G, and $\delta=4.2$.
    

\begin{references}

\reference {} Bahcall, J. N. et al. 1995, ApJ, 452, L91 

\reference {} Chartas, G. et al. 2000, ApJ, in press (astro-ph/0005227)

\reference {} Conway, R. G., Garrington, S. T., Perley, R. A., \&
Biretta, J. A. 1993, A\&A, 267, 347 

\reference {} Davis, R.J., Muxlow, T.W.B., \& Unwin, S.C. 1991,
Nature, 354, 374 

\reference {} Elvis, M., Wilkes, B.J., \& Lockman, F. 1989, AJ, 97,
777 

\reference {} Flatters, C. \& Conway, R.G. 1985, Nature, 314, 425

\reference {} Garmire, G. et al. 2000, ApJS, submitted 

\reference {} Gehrels, N. 1986, ApJ, 303, 336 

\reference {} Harris, D. E. \& Stern, C. P. 1987, ApJ, 313, 136 

%\reference {} Holtzman, J. A., Burrows, C. J., Casertano, S., Hester,
%J. J., Trauger, J. T., Watson, A. M., \& Worthey, G. 1995, PASP, 107,
%1065

\reference {} Mannheim, K., \& Biermann, P. L. 1992, A\&A, 253, L21

\reference {} Neumann, M., Meiseinheimer, K., \& R\"oser, H.-J. 1997,
A\&A, 326, 69

\reference {} R\"oser, H.-J., Meisenheimer, K., Neumann, M., Conway,
R. G., \& Perley, R. A. 2000, A\&A, 360, 99 

\reference {} R\"oser, H.-J., Meisenheimer, K., Neumann, M., Conway,
R. G., Davis, R.J., \& Perley, R. A. 1997, Reviews in Modern
Astronomy, 10, 253 

\reference {} R\"oser, H.-J., Conway, R. G., \& Meisenheimer, K. 1996,
A\&A, 314, 414 

\reference {} R\"oser, H.-J. \& Meisenheimer, K. 1991, A\&A, 252, 458 

\reference {} Tavecchio, F., Maraschi, L., Sambruna, R.M., \& Urry,
C.M. 2000, ApJL, submitted 

\reference {} Tavecchio, F., Maraschi, L., \& Ghisellini, G. 1998,
ApJ, 509, 608 

\reference {} von Montigny, C. et al. 1997, ApJ, 483, 161

\end{references}
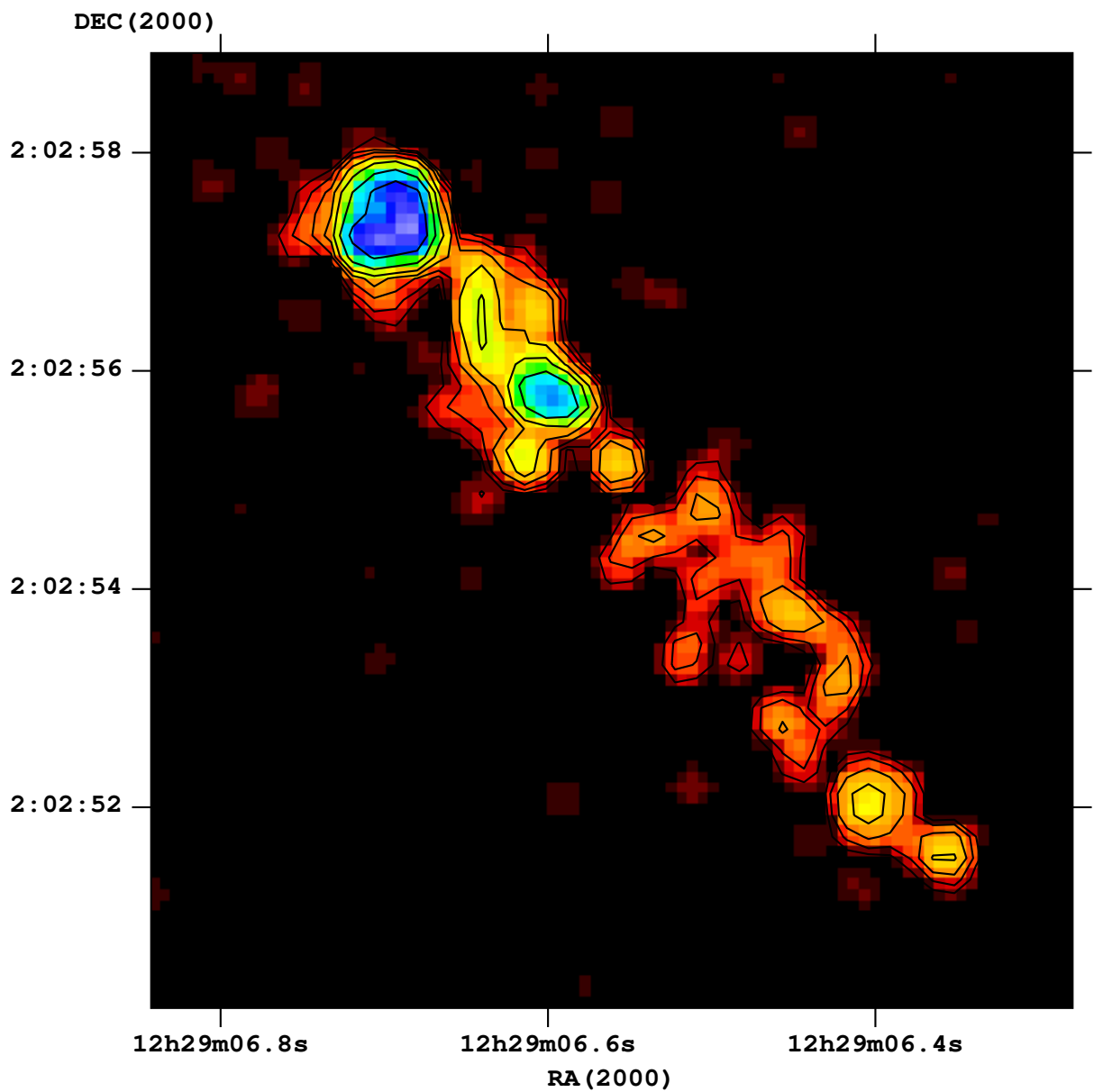
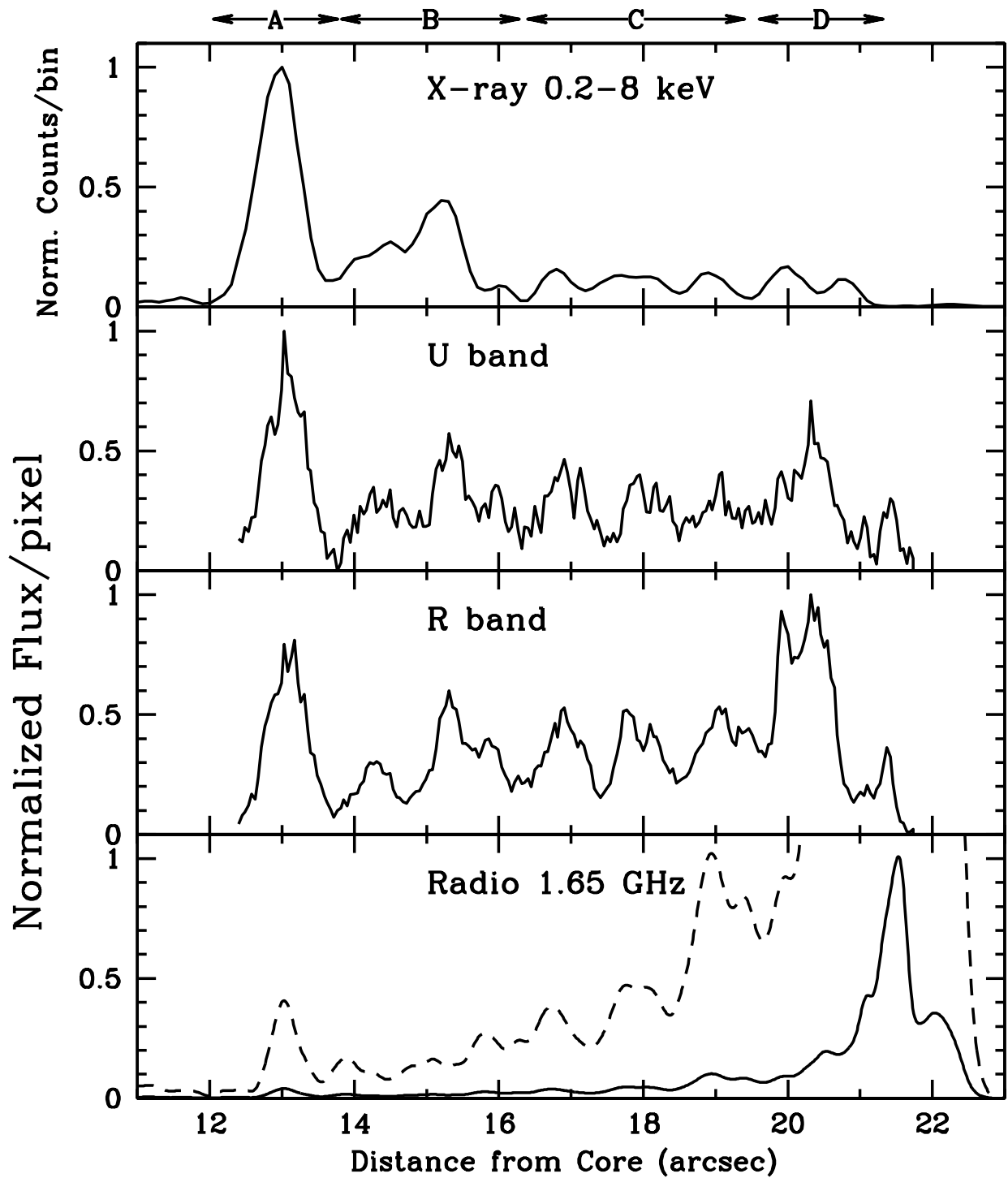
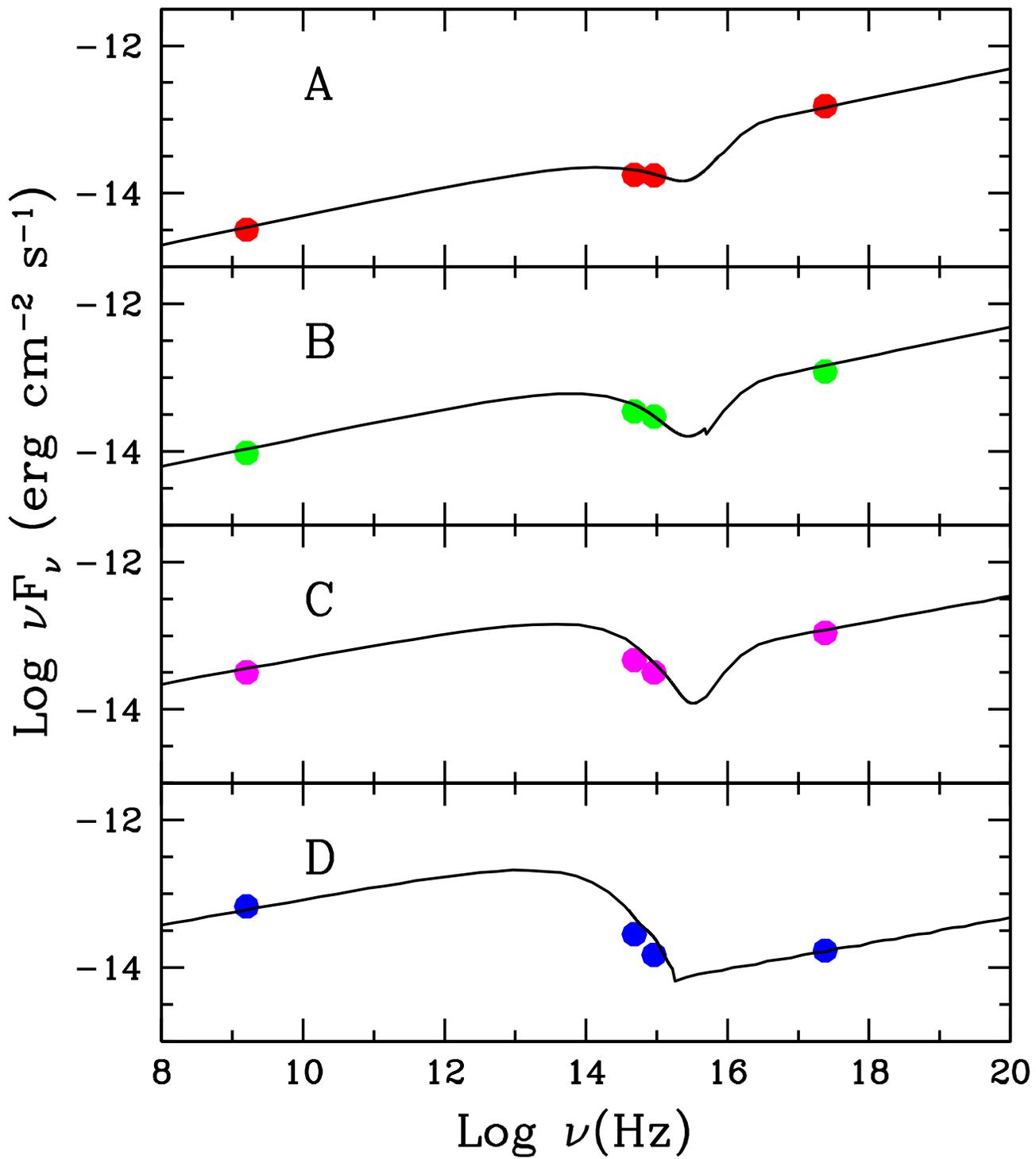
\end{document}